\documentclass[amsmath,amssymb,aps,eqsecnum]{revtex4}
\usepackage[dvips]{graphicx}

\begin{document}
\title{Statistical properties of random density matrices}
\author{Hans-J{\"u}rgen Sommers$^1$ and Karol {\.Z}yczkowski$^{2,3}$}

\affiliation{$^1$Fachbereich Physik,
Universit\"{a}t Duisburg-Essen, Campus Essen, 
  45117 Essen, Germany}

\affiliation {$^2$ Instytut Fizyki im. Smoluchowskiego,
Uniwersytet Jagiello{\'n}ski,
ul. Reymonta 4, 30-059 Krak{\'o}w, Poland}
\affiliation{$^3$Centrum Fizyki Teoretycznej, Polska Akademia Nauk,
Al. Lotnik{\'o}w 32/44, 02-668 Warszawa, Poland}

 \date{July 16, 2004}

\begin{abstract}
Statistical properties of ensembles of random density matrices
are investigated. We compute  traces and von Neumann
entropies averaged over ensembles of random density matrices
distributed according to the Bures measure.
The eigenvalues of the random density matrices are analyzed: we derive 
the eigenvalue distribution for the Bures ensemble which is shown to be broader
then the 
quarter--circle distribution characteristic of the Hilbert--Schmidt ensemble.
For measures induced by partial tracing over the environment we compute exactly
the
two-point eigenvalue correlation function.   
\end{abstract}


\maketitle

\medskip
\begin{center}
{\small e-mail: sommers@next30.theo-phys.uni-essen.de
  \ \quad \ karol@cft.edu.pl}
\end{center}


\section{Introduction}

Analyzing the density matrices of a finite size $N$ one is often 
interested in the properties of typical states. In general 
the properties depend on the measure, according to which
the random density matrices are distributed. Nowadays it is widely accepted,
that it is not possible to single out the only,
unique measure in the set ${\cal M}_N$ of all density matrices of size $N$. 

However, several possible measures are distinguished by different
mathematical and physical arguments. For instance the 
Hilbert-Schmidt (HS) measure arises if one constructs 
random pure states $|\Psi\rangle$ distributed
 according to the natural (Fubini--Study)
measure on the space of pure states on a composite Hilbert space
${\cal H}_N \otimes {\cal H}_N$ and obtains mixed states
by partial tracing, $\rho={\rm Tr}_N(|\Psi\rangle\langle \Psi|)$
\cite{ZS01}. Moreover, the HS measure may be defined 
by the HS distance,
\begin{equation}
D_{\rm HS}(\rho,\sigma)=\sqrt{{\rm Tr} [(\rho - \sigma)^2]},
\label{HS}
\end{equation}
which induces the flat geometry in ${\cal M}_N$.
For instance, the set of $N=2$ mixed states 
analyzed with respect to the HS distance
displays the geometry of the $3$--ball (the Bloch ball),
with the Bloch sphere, containing the pure states, at its boundary. 

Another measure in the space of mixed states, which should be distinguished, is 
the Bures measure \cite{Ha98}. It is induced by the Bures metric
\cite{Bu69,Uh92}
\begin{equation}
D_B(\rho,\sigma)=
\sqrt{2}\bigl[ 1- {\rm Tr}(\sqrt{\rho}\sigma \sqrt{\rho})^{1/2}\bigr] ^{1/2} .
\label{bures}
\end{equation}
which is  Riemannian and monotone.
It is a Fisher--adjusted metric \cite{PS96},
since in the subspace of diagonal matrices it induces
the statistical distance \cite{BC94}. Moreover, the Bures metric is
Fubini--Study adjusted, since at the space of pure states 
both metrics do agree \cite{Uh95}.
These unique features of the Bures distance are used
to support the claim
that without any prior knowledge on a certain density matrix,
the optimal way to mimic it is to generate it at
random with respect to the Bures measure. 

In this work we analyze statistical properties 
of ensembles of random states. In section II we
provide the definitions of the Hilbert--Schmidt 
and the Bures ensembles of random density matrices and recall
their joint distribution functions for their spectra. The expectation values 
of the moments $\langle {\rm Tr} \rho^q\rangle $
and von Neumann entropies are computed in section III. 
In section IV we analyze the eigenvalue density of random 
states: the quarter--circle distribution characteristic of the
Hilbert--Schmidt ensemble is rederived and compared with an 
explicit distribution computed for the Bures ensemble.
Section V is devoted to the ensembles of random matrices
obtained by partial tracing, for which the average traces
and the eigenvalue correlation functions are computed.

\section{Ensembles of random states}

We are concerned with ensembles of random states, for which the probability
measure has a product form and may be factorized \cite{Ha98,ZS01}, 
\begin{equation}
{\rm d} \mu_{\rm x} = {\rm d} \nu_{\rm x} 
 (\lambda_1,\lambda_2,...,\lambda_N) \times {\rm d} h.
  \label{product}
\end{equation}
The latter factor, d$h$, determining the distribution of the eigenvectors of 
the density matrix, is the unique, unitarily invariant, Haar measure on
$U(N)$. On the other hand, the first factor describing the
distribution of eigenvalues $\lambda_i$ of $\rho$ depends on the measure 
used (the label $_{\rm x}$ denotes any of the product measures investigated).

The Hilbert-Schmidt measure induces the following joint 
distribution function 
in the simplex of eigenvalues \cite{Ha98,ZS01}
\begin{equation}
 P_{\rm HS}({\vec \lambda}) = 
 \frac{\Gamma(N^2)}
{\prod_{j=0}^{N-1} \Gamma(N-j) \Gamma(N-j+1) }
\ \delta\Bigl(1-\sum_{j=1}^N \lambda_j \Bigr)  
\ \prod_{i<j}^N (\lambda_i-\lambda_j)^2\ .
  \label{HSmes}
\end{equation}
This distribution may be considered as a special case of the
family of measures induced induced by partial tracing.
Consider a pure state of  
$|\Phi\rangle \in {\cal H}_N \otimes {\cal H}_K$
of a composite bi--partite system of size $NK$. 
Tracing over the $K$--dimensional environment
one obtains a mixed state of size $N$,
namely $\rho={\rm T}_K\bigl(|\Phi\rangle \langle \Phi|\bigr)$.
A natural assumption that 
$|\Phi\rangle$ is a random pure state
distributed according to the unique, unitarily
invariant measure on the set of pure states
leads to the family of measures 
\begin{equation}
 P_{{N,K}}({\vec \lambda}) =
\frac{\Gamma(KN)}
{\prod_{j=0}^{N-1} \Gamma(K-j) \Gamma(N-j+1) }
\delta\Bigl(1-\sum_{j=1}^N \lambda_j \Bigr)
\  \prod_i\lambda_i^{K-N}
\prod_{i<j}(\lambda_i-\lambda_j)^2 \ ,
  \label{HSindNK}
\end{equation}
labeled by the size $K$ of the environment. 
Such induced measures were discussed many times 
in the literature \cite{Lu78,LS88,Pa93},
while the normalisation constant was derived in 
\cite{ZS01}. Note that in the symmetric 
case $K=N$ the induced measure 
reduces to the Hilbert-Schmidt  measure (\ref{HSmes}).

It is worth to emphasize a link to known ensembles of random matrices.
In order to construct a random density matrix according to the measure
$\mu_{N,K}$ it is sufficient to generate a rectangular Gaussian 
matrix $X$ of size $N \times K$ 
and to compute $\rho=X^{\dagger}X/{\rm Tr}( X^{\dagger}X)$ \cite{ZS01}.
In the special case $K=N$ this fact shows a 
relation between the Ginibre ensemble of non Hermitian random matrices
\cite{Me91} and the Hilbert-Schmidt measure.

The Bures measure in the simplex of eigenvalues 
may be derived from an assumption that any ball
in the sense of the Bures distance of a fixed radius
belonging to the set  ${\cal M}_N$ 
has the same volume. The Bures probability
distribution in the simplex of eigenvalues 
was obtained by Hall \cite{Ha98}
\begin{equation}
 P_{\rm B}({\vec \lambda}) =
C_N \frac{\delta\bigl(\lambda_1+\lambda_2+...+\lambda_N - 1)}
          {\sqrt{\lambda_1\lambda_2 \cdots \lambda_N}}
  \ \prod_{i<j}
\frac{ (\lambda_i-\lambda_j)^2 } 
     {\lambda_j+\lambda_j} .
  \label{mesbur2}
\end{equation}
The normalization constants $C_N$ were found by Slater \cite{Sl99b} 
for low values of $N$, while the general formula
\begin{equation}
  C_N=2^{N^2-N}\ \frac{\Gamma(N^2/2)}
  {\pi^{N/2}\  \prod_{j=1}^{N} \Gamma(j+1) }
 \label{Hallconst}\ 
\end{equation}
was derived in \cite{SZ03}. The volume of the set
of mixed quantum states and the area of its boundary
were computed with respect to both measures in
\cite{ZS03,SZ03}.

\section{Expectation values}

To characterize, to what extend a given state $\rho$ is mixed
one may use the moments, Tr$\rho^q$, with any $q> 0$.
The simplest to compute is the second trace
$r={\rm Tr}\rho^2$, called {\sl purity}, which is closely related
to the {\sl linear entropy} $1-r$ and inverse participation ratio, 
$R=1/r$. Mean purity averaged over the HS measure
is smaller then the average over the Bures measure, 
 \begin{equation}
  \langle {\rm Tr} \rho^2 \rangle_{\rm HS} =
\frac{2N}{N^2+1} \quad < \quad
\langle {\rm Tr} \rho^2 \rangle_{\rm B} =
\frac{5N^2+1}{2N(N^2+2)}.
 \label{trace2}
 \end{equation}
This result reflects the fact 
that the Bures  measure is more concentrated on the states of 
higher purity, than the Hilbert-Schmidt measure \cite{ZS01}.
It is not so simple to get such results for an arbitrary exponent $q$.
However, it is easier to perform averaging in the 
asymptotic regime, $N>> 1$.
Mean traces, averaged  over
the Hilbert-Schmidt measure are 
 \begin{equation}
  \langle {\rm Tr} \rho^q \rangle_{\rm HS} =
   N^{1-q} \frac{\Gamma(1+2q)}{\Gamma (1+q) \Gamma(2+q)}
  \left( 1+O\left({1\over N}\right) \right)\ .
 \label{traceHS}
 \end{equation}
The analogous average with respect to the Bures measure reads 
 \begin{equation}
  \langle {\rm Tr} \rho^q \rangle_{\rm B} =
   N^{1-q} 2^q  \frac{ \Gamma[(3q+1)/2] }
                  {\Gamma [(1+q)/2] \Gamma(2+q)}
  \left( 1+O\left({1\over N}\right) \right)\ .
 \label{traceB}
 \end{equation}
Again we find $\langle {\rm Tr} \rho^q \rangle_{\rm HS} <\langle {\rm Tr}
\rho^q \rangle_{\rm B}$.
As a measure of the degree of mixing one often uses
the von Neumann entropy, $S(\rho ) = -\mbox{Tr} \rho \ln{}\rho$.
It varies from $S(\rho)=0$ for any pure state and
  $S(\rho)=\ln N$ for the maximally mixed state.
Since $S(\rho) =-\lim_{q\to 1} \partial {\rm Tr} \rho^q / \partial q$
the mean von Neumann entropy may be obtained by differentiation
of (\ref{traceHS}) and (\ref{traceB}) with respect to the
parameter,
$\langle S \rangle 
=  -\lim_{q\to 1} \langle \partial {\rm Tr} \rho^q / \partial q \rangle
=-\lim_{q\to 1} \partial
\langle {\rm Tr} \rho^q \rangle  / \partial q$.
The results are
 \begin{equation}
  \langle S(\rho) \rangle_{\rm HS} =
   \ln N - \frac{1}{2} +
   O\left( \frac{\ln N}{N} \right)  .
\label{entrHS}
 \end{equation}
for the Hilbert-Schmidt measure and 
 \begin{equation}
  \langle S(\rho) \rangle_{\rm B} =
   \ln N - \ln 2 +
   O\left( \frac{\ln N}{N} \right)  .
 \end{equation}
 for the Bures measure.
Note that the former result is larger,
since the HS measure favors more mixed states.
Although the mean value of the traces
which respect to the HS measure appeared several times
in the literature \cite{Lu78,LS88,Pa93}, 
the results for the Bures measure are new.
Their derivation is sketched in  appendix A. 
Using the expansion of the generating functions there, it is
possible to give some more moments for the Bures measure. 
We compare them with the previously  known Hilbert--Schmidt averages 
\begin{equation}
\langle {\rm Tr}\rho^3\rangle_{\rm HS} = {5N^2+1 \over (N^2+1)(N^2+2)},\ \ \ \
\
\ \ \ \ \ \langle {\rm Tr}\rho^3\rangle_{\rm B}= {8N^2+7 \over (N^2+2)(N^2+4)}
, 
\end{equation}
\begin{equation}
\langle {\rm Tr}\rho^4\rangle_{\rm HS} = {14N^3+10N \over
(N^2+1)(N^2+2)(N^2+3)},\ \ \ \ \ \ \ \ \ \ 
\langle {\rm Tr}\rho^4\rangle_{\rm B}=
{21(11N^4 +25 N^2 +4) \over 8N(N^2+2)(N^2+4)(N^2+6)}.
\end{equation}

\section{Distribution of eigenvalues}

We are going to evaluate the distribution of the rescaled eigenvalue
$x:=N\lambda_1$ in the limit of large dimension  $N$ of density matrices.
To derive the probability distribution $P(x)$, we analyze
$q$-th moments of these distributions. For the  
Hilbert--Schmidt  measure we obtain
\begin{equation}
f_{\rm HS}(q)=\int P_{\rm HS}(x) x^q {\rm d} x= \frac{1}{\pi}
2^{2q} \frac{\Gamma(q+1/2) \Gamma(1/2)}
            {\Gamma(q+2)}  ,
\label{momntHS}
\end{equation}
while the moments for the Bures measure are
\begin{equation}
f_{\rm B}(q)=\int P_{\rm B}(x) x^q {\rm d} x= \frac{1}{2\pi}
2^{q}3^{3q/2}  \frac{ \Gamma (q/2+1/6)
\Gamma (q/2+5/6)}
{\Gamma (q+2)}\ .  
\label{momntB}
\end{equation}
The above results follow from eqs.(\ref{traceHS}, \ref{traceB})
by use of the duplication and triplication formula for the Gamma function.
They allow us to obtain the explicit form of the 
level density, exact in the asymptotic limit of large $N$. 
The distribution obtained for the HS measure 
\begin{equation}
P_{\rm HS}(x)=  \frac{1}{2\pi}
 \sqrt{\frac{4}{x} -1}
{\quad \rm for \quad}
x\in [0,4]
\label{densitHS}
\end{equation}
diverges as $x^{-1/2}$ for $x\to 0$ and 
becomes a quarter--circle law in
the rescaled variable, $y=\sqrt{x}$ - see Fig. 1.
It is comforting to verify that this law 
forms a special case of the distribution
obtained  by Page  for the induced measures \cite{Pa93}
and later derived in a different context in \cite{SM99}.  
On the other hand, the distribution for the Bures measure
\begin{equation}
P_{\rm B}(x)=  \frac{3}{4a\pi} \left[ 
\left(\frac{a}{x} +\sqrt{\left({{a}\over {x}}\right)^2-1} \right)^{2/3} -
\left(\frac{a}{x} -\sqrt{\left({{a}\over {x}}\right)^2-1} \right)^{2/3}\
\right]
{\quad \rm for \quad}
x\in (0,a]
\label{densitB}
\end{equation}
is defined on a larger support,
$x\le a=3\sqrt{3}$, and
diverges for $x\to 0$  as $x^{-2/3}$.
Level repulsion for the Bures ensemble compared to the HS ensemble will be
reduced at $x=0$ but enhanced at the maximum of the spectrum.
 
\begin{figure} 
   \begin{center}
 \vskip -0.2cm
\includegraphics[width=12.0cm,angle=0]{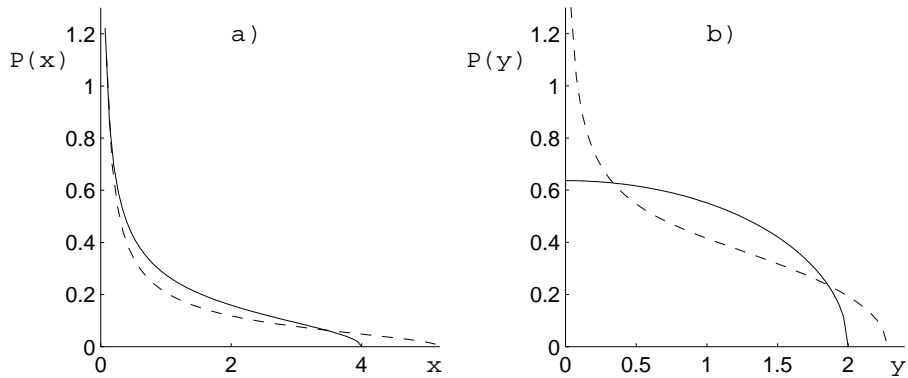}
\vskip -0.2cm
\caption{Level density of random density matrices $P(x)$
for Bures measure (dashed line) and Hilbert--Schmidt measure (solid line) a).
The latter becomes a quarter-circle distribution  in the rescaled
variable, $y=\sqrt{x}$ b).}
 \label{fig:PxHSB}
\end{center}
 \end{figure}

The above distributions may be derived in an alternative way
by minimization of the action functional
\begin{equation}
A_{\rm HS} = - \int {\rm d} x {\rm  d}x' P(x) P(x') \ln |x-x'| 
\label{actionHS}
\end{equation}
for the HS measure, and
\begin{equation}
A_{\rm B} = A_{\rm HS} +
  \frac{1}{2} \int {\rm d} x {\rm  d}x' P(x) P(x') \ln (x+x') \ .
 \label{actionB}
\end{equation}
for the Bures measure.
Both (unknown) solutions of these minimization problems 
should satisfy the normalization condition,
$\int P(x) {\rm d}x=1$ and the relation induced by the unit trace constraint,
$\int x P(x) {\rm d}x=1$. Both conditions can be implemented with the help of
Lagrange multipliers. 
The resulting  integral equations for $P(x)$ may be solved by the Green
function
\begin{equation}
G(t) = \int {\rm d}x \frac{P(x)}{x-t}
 \label{Green}
\end{equation}
where the cut along the real axis gives the densities (\ref{densitHS},
\ref{densitB}).
The Hilbert-Schmidt measure leads to a rather simple Green function
\begin{equation}
G_{\rm HS}(t) = \frac{1}{2} \left( \sqrt{1-{4\over t}} - 1 \right), 
 \label{GreenHS}
\end{equation}
for $t<0$ and otherwise given by analytic continuation,
  corresponding to (\ref{densitHS}). 
The Green function corresponding to the Bures measure is 
more complicated
\begin{equation}
G_{\rm B}(t) = \frac{1}{6} \left( z+\frac{1}{z} -1\right)
{\rm \quad with \quad}
z=\left(-\frac{a}{t}\right)^{2/3} \ 
\left(1- \sqrt{1-\left({{t}\over {a}}\right)^2} \right)^{2/3}
 \label{GreenB}
\end{equation}
for $-a<t<0$ (otherwise given by analytic continuation)
and leads to the distribution (\ref{densitB}). 
The Green functions fulfill the generalized {\sl Pastur equations}
\cite{pastur}
\begin{equation}
G_{\rm HS}(t) = \frac{-1/t} {1+G_{\rm HS}(t)}
{\rm \quad and \quad}
G_{\rm B}(t) = \frac{-1/t} {\sqrt{1+2G_{\rm B}(t)}} ,
 \label{Pastur}
\end{equation}
which suggests an interpolation formula between the Hilbert--Schmidt measure,
($\alpha=1$), and
the Bures measure, ($\alpha=2$) :
\begin{equation}
G_{\alpha}(t) = \frac{-1/t} {[1+\alpha G_{\alpha}(t)]^{1/\alpha}}\ . 
 \label{Pastur2}
\end{equation}
It would be interesting to analyze  the family of interpolating measures which
lead to the
above Green functions. 

\section{Eigenvalue density and eigenvalue correlation for induced measures}

In this section we are going to investigate statistical 
properties of the induced measures (\ref{HSindNK})
defined in the space ${\cal M}_N$ 
of density matrices of size $N$. An integer $K\ge N$,
represents the size of an environment and may be treated as a 
parameter labeling the measure.

\subsection{Eigenvalue density}

The one-point density
$P(\lambda)$ is obtained from the Green function 
\begin{equation}
 G(\lambda)= \left \langle {1\over N}\sum_i{1\over \lambda_i-\lambda}\right
\rangle_{N,K}\ \ \ {\rm with} \ \ \ P(\lambda)={1\over \pi} {\rm
Im}G(\lambda+i\delta)\ , 
\label{GreenF}
\end{equation}
and the Green function will be derived from the generating function
\begin{equation}
 Z(\mu)= \left \langle \prod_i\left({\lambda_i-\mu \over
\lambda_i-\lambda}\right)\right \rangle_{N,K}\ \ \ {\rm with} \ \ \
G(\lambda)=-{1\over N}{{\partial\over \partial\mu}} Z|_{\mu=\lambda}\ . 
\label{GenF}
\end{equation}
Due to the structure of $P_{N,K}$ with the van der Monde determinant we may
write
$Z(\mu)$ as inverse Laplace transform of a determinantal function:
\begin{equation}
Z(\mu) \propto \int_{-i\infty+\epsilon}^{+i\infty+\epsilon} {ds\over 2\pi i}
e^s \det\left(\int_0^{\infty}dx {\rm e}^{-sx}x^{K-N+i+j-2}\left({x-\mu\over
x-\lambda}\right)\right)  
\label{Zmu}
\end{equation}
with $i,j=1,2,...,N$. From this we immediately obtain
\begin{equation}
 G(\lambda)={\Gamma(KN)\over N} \int_{-i\infty+\epsilon}^{+i\infty+\epsilon}
{ds\over 2\pi i} {{\rm e}^s\over s^{KN}}
\sum_{i,j=1}^N W^{-1}_{j,i}\int_0^{\infty}dx{{\rm e}^{-sx} x^{K-N+i+j-2}\over
(x-\lambda )s^{K-N+i+j-1}}\ .  
\label{Gla}
\end{equation}
Here the matrix $W$ of size $N$ is given by 
\begin{equation}
 W_{i,j} = \Gamma(K-N+i+j-1)\ .   
\label{Wij}
\end{equation}
An explicit form of the matrix $W^{-1}$ reads 
\begin{equation}
 (W^{-1})_{i,j} = (-1)^{i+j}\sum_{k={\rm max}(i,j)}^N 
{k-1\choose i-1}{k-1 \choose
j-1}{\Gamma(K-N+k)\over \Gamma(k)\Gamma(K-N+i) \Gamma(K-N+j) }\ .  
\label{W-1ij}
\end{equation}
It turns out that the $x$ integral in Eq. (\ref{Gla}) contributes only for
$x<1$ and we obtain as result
\begin{equation}
 G(\lambda)={\Gamma(KN)\over N}  \sum_{i,j=1}^N W^{-1}_{j,i}\int_0^{1}dx
{x^{K-N+i+j-2}(1-x)^{KN-(K-N+i+j)}\over \Gamma[KN-(K-N+i+j-1)](x-\lambda)} \ .  
\label{Gla1}
\end{equation}
The normalization constant in eqs. (\ref{Gla}, \ref{Gla1}) has been restored by
the known asymptotic behavior of $G(\lambda)$. Thus we immediately obtain the
density
\begin{equation}
 P(\lambda)={\Gamma(KN)\over N}  \sum_{i,j=1}^N W^{-1}_{j,i} 
{\lambda^{K-N+i+j-2}(1-\lambda)^{KN-(K-N+i+j)}\over \Gamma[KN-(K-N+i+j-1)] } \
.  
\label{rho1}
\end{equation}
The above rather complicated form of obtaining the density
already derived by Page \cite{Pa93},
allows us to calculate the moments 
with the help of Euler's Beta-function  
\begin{equation}
 \langle \lambda^q \rangle  ={\Gamma(KN)\over N\Gamma(KN+q)} 
\sum_{i,j=1}^N W^{-1}_{j,i}  \Gamma(K-N+i+j-1+q)  \ .  
\label{moments}
\end{equation}
From this follows the mean von Neumann entropy
\begin{equation}
 \langle S \rangle =-N\langle \lambda\ln \lambda \rangle = \psi(KN+1)-{1\over
KN }  \sum_{i,j}W^{-1}_{j,i}  \Gamma(K-N+i+j ) \psi(K-N+i+j )   \ .  
\label{moments1}
\end{equation}
where $\psi(x)$ is Euler's Digamma-function = $\Gamma'(x)/\Gamma(x)$, and where
we have suppressed the index $_{N,K}$ at the angular brackets. 
The result for the average entropy has been conjectured
by Page \cite{Pa93}, and 
later proved in \cite{FK94,SR95,Se96}. It is a rational number.
All the formulas are valid for $K\ge N$. Again for $K<N$ one has to interchange
$K$ and $N$, obtaining density and moments of the $K$ positive eigenvalues.

One may explicitly give some average traces
over the induced measure $\mu_{N,K}$
\begin{equation}
\langle {\rm Tr}\rho^2\rangle = {K+N\over KN+1} , 
\quad \quad
\langle {\rm Tr}\rho^3\rangle= {(K+N)^2 +KN +1 \over (KN+1)(KN+2)}, 
\end{equation}
\begin{equation}
\langle {\rm Tr}\rho^4\rangle= { (K+N)[(K+N)^2 +3KN +5] \over
(KN+1)(KN+2)(KN+3)},
\end{equation}
the first of which appeared already in the paper of Lubkin \cite{Lu78}, 
the others are consistent with the recent work of Malacarne et al.
\cite{MML02}.
 Here
$\rho$ means again the density matrix and not the eigenvalue density. To find
the coefficients of the polynomial in the denominator it is useful to know its
order and symmetry as can be found going back to a Gaussian integral writing
the density matrix as a matrix of Wishart form $\rho =\psi \psi^\dagger$. In
principle they are contained in formula (\ref{moments}).

\subsection{ Eigenvalue correlation}

The eigenvalue correlation can be obtained from the Green-function correlation
\begin{equation}
  \left\langle {1\over N}\sum_i{1\over \lambda_i-\lambda}{1\over
N}\sum_i{1\over \lambda_i-\mu}\right\rangle \ , 
\end{equation}
 which can be derived from
the generating function
\begin{equation}
\left\langle \prod_i\left({(\lambda_i-\kappa_1)(\lambda_i-\kappa_2) \over
(\lambda_i-\lambda)\ (\lambda_i-\mu)}\right)\right\rangle\ . 
 \end{equation}
The result for
the two-eigenvalue density $P(\lambda,\mu)$, which can be obtained along
the same lines as in subsection A is then:
\begin{equation}
 P(\lambda,\mu)= \theta(1-\lambda-\mu){\Gamma(KN)\over N(N-1)} 
\sum_{i,j,k,l=1}^N [W^{-1}_{j,i}W^{-1}_{l,k}-W^{-1}_{l,i}W^{-1}_{j,k}] 
{\lambda^{K-N+i+j-2}\mu^{K-N+k+l-2}(1-\lambda-\mu)^{KN-2K+2N-i-j-k-l+1}\over
\Gamma(KN -2K+2N-i-j-k-l+2) } \ .  
\label{rholm}
\end{equation}
We have checked: $P(\lambda)=\int P(\lambda, \mu)\ d\mu$.
In order to prove this, scale $\mu$ with $1-\lambda$, integrate over $\mu$
with the help of Euler's Beta--function, and use eq. (5.5). Then the factor
$N-1$ in the denominator of eq. (5.15) cancels and the result is eq. (5.8).
The first bracket under the sum in (5.15) ensures level repulsion $\propto
(\lambda-\mu)^2$ for $\lambda-\mu \to 0$.
Furthermore there is additional repulsion from the
boundaries at $ \lambda=0, \mu=0, 1-\lambda-\mu=0$.

 It is again easy to calculate the moments with the help of Euler's
Beta--function
\begin{equation}
 \langle \lambda^L \mu^M \rangle  ={\Gamma(KN)\over N(N-1)\Gamma(KN+L+M)} 
\sum_{i,j,k,l=1}^N [W^{-1}_{j,i}W^{-1}_{l,k}-W^{-1}_{l,i}W^{-1}_{j,k}] 
\Gamma(K-N+i+j-1+L)\Gamma(K-N+k+l-1+M)  \ .  
\label{moments2}
\end{equation}
For the entropy correlation we have
\begin{equation}
  \langle S\ S \rangle = N(N-1)\langle \lambda(\ln \lambda)\ \mu(\ln \mu) 
\rangle +\ N\langle \lambda^2(\ln \lambda)^2 \rangle \ ,  
\label{entropyc}
\end{equation}
which can be obtained by double differentiation of $\langle \lambda^L \mu^M
\rangle$ and $\langle \lambda^{L+M} \rangle$ with respect to $L$ and $M$ at
$L=M=1$.
Again one may obtain formulas for $K<N$ by interchange of $K$ and $N$.

\section{Concluding remarks}

It is well known that that there is no single,
naturally distingushed probability measure
in the set of mixed quantum states of a fixed
size $N$. Guessing a mixed state on random 
without any additional information whatsoever,
it is legitimate to use the Bures measure (\ref{mesbur2}),
related to the statistical distance and distinguishability.
On the other hand, if it is known
that the mixed state has arisen by the partial tracing
over a $K$ dimensional environment, 
one uses the induced measure (\ref{HSindNK}),
which reduces to the Hilbert-Schmidt
measure in the special case $K=N$.

In this work we investigated statistical properties
of ensembles of density matrices of a fixed size 
generated according to the Bures or the 
Hilbert--Schmidt measure. We computed the averages
over the set of mixed quantum states with respect to both measures
and derived the level density in the asymptotic limit of large matrices.
Furthermore, for measures obtained from random pure
 states of a composite system by
partial tracing we computed the one--point
eigenvalue density, the exact two--point eigenvalue density, the
corresponding moments and average entropies.
On one hand, results concerning average traces and 
average entropy may be useful from the point of view of the theory of
quantum information \cite{NC00}: the von Neumann entropy of a mixed state
$\rho$
is equal to the entanglement of the pure state $|\Psi\rangle$ belonging to a
composed Hilbert space, which purifies $\rho$. On the other hand, results
obtained
contribute to the theory of random matrices: the ensembles of random density
matrices distributed according to the Bures measure
display different properties then the standard Gaussian
ensembles of Wigner and Dyson \cite{Me91}.

It is worth to add that the ensembles of random states analyzed in this work
do not cover all the cases of a physical intrest. For instance it is 
natural to assume that in a concrete experiment a mixed state $\rho$
is formed by applying a known quantum channel $\Phi$
(completely positive, trace preserving map) on a random pure state,
\begin{equation}
\rho=\Phi(|\psi\rangle\langle \psi|).
\label{ensemoper}
\end{equation}
Without any information concerning the pure state $|\psi\rangle\in {\cal H}_N$
one has to assume that it is generated according to the natural, Fubini--Study
 measure in the set of pure states. In this manner 
 any quantum channel $\Phi$ induces 
by (\ref{ensemoper}) a certain measure in the space of mixed quantum states.
Hence it would be interesting to repeat the computation 
performed in this work for ensembles of mixed states
obtained by physically motivated quantum channels. 
Such a research will be a subject of
a forthcoming publication.


It is a pleasure to thank  D. Savin for fruitful discussions.
Financial support by the Sonderforschungsbereich /Transregio 12
der Deutschen Forschungsgemeinschaft 
and Komitet Bada{\'n} Naukowych in Warsaw under
the grant 1~P03B~042~26 is gratefully acknowledged.


\appendix
\section{Moments for Bures measure }

One may derive all moments for the Bures distribution $P_B(\rho)$
from a Laguerre type ensemble
\begin{equation}
  P_B^L(\rho) \propto \theta(\rho) {\rm e}^{-{\rm Tr}\rho}\
\prod_{i,j}^{1...N}(\lambda_i+\lambda_j)^{-1/2} \ ,  
\label{PLB}
\end{equation}
where $\lambda_i$ denote eigenvalues of $\rho$.
Moments are related by
\begin{equation}
  \int M_p(\rho)P_B(\rho)D\rho = {\Gamma(N^2/2)\over \Gamma(N^2/2+p)} \int
M_p(\rho)P_B^L(\rho)D\rho\\ ,  
\label{Mprho}
\end{equation}
where $M_p(\rho)$ is a homogeneous function of $\rho$ of degree $p$ and $D\rho$
is the matrix volume element of a Hermitian matrix. Next we may write
\begin{equation}
P_B^L(\rho) \propto \theta(\rho) \int DA\ {\rm e}^{-{\rm Tr}[\rho(1+A^2)]}     
\label{PBL1}
\end{equation}
with a Hermitian matrix $A$.
Let us denote its eigenvalues by $\{A_i\}$.
 In the following we use the formula \cite{SZ03}
\begin{equation}
  \theta(\rho)   {\rm e}^{-{\rm Tr}(\rho\epsilon)} = B_N \det(\delta/\delta\rho
+\epsilon)^{-N}\ \delta(\rho)   
\label{Form}
\end{equation}
with   a positive definite Hermitian matrix $\epsilon$ and
\begin{equation}
  B_N=\pi^{N(N-1)/2} \Gamma(1)\Gamma(2)...\Gamma(N)\ .  
\label{BN}
\end{equation}
Note that on the left hand side of Eq. (\ref{Form}) there is the restriction
$\rho
\ge 0$, while on the right hand side we have no restriction on integration for
$\rho$.
With the above formula it is easy to compute the matrix Laplace transform
\begin{equation}
 \int {\rm e}^{-{\rm Tr} (E\rho)}P_B^L(\rho)D\rho \propto \int DA\ 
\det(E+1+A^2)^{-N} 
\label{MLaplace}
\end{equation}
with a nonnegative matrix $E$ of size $N$. 
For $E=0$ this formula leads to the normalization
constant for the Bures measure \cite{SZ03}. 
Let $\{E_1,\dots,E_N\}$ denote the eigenvalues of $E$. 
With the help of the
Itzykson-Zuber integral \cite{itzykson,harish} the right hand side of Eq. 
(\ref{MLaplace})
 is proportional to 
 \begin{equation}  
\int dA_1 ...dA_N  {[\Delta(A)]^2\over \Delta(A^2)\Delta(-E)} 
\det\bigl[ \frac{1}{1+A_i^2+E_j} \bigr] 
\end{equation}
with the van der Monde determinant $\Delta(A)=\prod_{i<j}(A_i-A_j)$.
Finally one may perform the $A_i$ integrations in the complex plane, arriving
at the generating function
\begin{equation}
  Z_B^L(E)=\int {\rm e}^{-{\rm Tr}
(E\rho)}P_B^L(\rho)D\rho=\prod_{i,j}^{1...N}{2\over
\sqrt{1+E_i}+\sqrt{1+E_j}}\ .
\label{ZLB}
\end{equation}
This expression has a rather simple expansion in powers of $E$.  It starts like
\begin{equation}
  Z_B^L(E)=1-{N\over 2}\sum E_i +\Bigl( {N^2\over 8} +{1\over 16}\Bigr)
\bigl( \sum E_i\bigr)^2 +
{3N\over 16}\sum E_i^2 + O(E^3) . 
\label{ZLBe}
\end{equation}
Thus it is possible to obtain all moments by matrix derivation,e.g.
\begin{equation}
  \langle {\rm Tr} F(\rho) \rangle_B^L= {\rm Tr} F(-\delta/\delta
E)Z_B^L(E)|_{E=0}\ .
\label{moments3}
\end{equation}
The corresponding generating function for the Hilbert-Schmidt measure is even
more simple and reads:
\begin{equation}
  Z_{HS}^L(E)=\int {\rm e}^{-{\rm Tr}
(E\rho)}P_{HS}^L(\rho)D\rho=\prod_{i=1}^{N}{1\over  (1+E_i)^N}\ .
\label{ZLHS}
\end{equation}
Its expansion starts like
\begin{equation}
  Z_{HS}^L(E)= 1-{N }\sum E_i +{N^2\over 2} \bigl( \sum E_i \bigr)^2 + {N\over
2}\sum E_i^2
+ O(E^3)\ .
\label{ZLHSe}
\end{equation}
To obtain the matrix derivatives $\delta/\delta E$ is not so easy in general,
since everything is expressed in eigenvalues $E_i$ e.g.
\begin{equation}
  \langle {\rm Tr} \rho^q \rangle_B^L= {\rm Tr} (-\delta/\delta
E)^qZ_B^L(E)|_{E=0} ={1\over \Delta(E) }\sum_{i=1}^N(-\partial/\partial
E_i)^q[\Delta(E)Z_B^L(E)]|_{E=0}\ .
\label{moments4}
\end{equation}
One can proof the last equation with the help of the Itzykson--Zuber integral.
The above formulae were used to derive the few results for Bures moments in
section III.


\begin{thebibliography}{99}

\bibitem{ZS01} {\.Z}yczkowski K and Sommers H-J  2001
 {\sl J. Phys.} {\bf A 34} 7111

\bibitem{Ha98}  Hall M J W 1998 {\sl Phys. Lett.}
{\bf A 242} 123

\bibitem{Bu69}  Bures  D J C 1969 {\sl Trans. Am. Math. Soc.}
  {\bf 135} 199

\bibitem{Uh92}  Uhlmann A  1992
in {\sl Groups and related Topics}  Gielerak R et. al.
(eds.) (Dodrecht: Kluver)


\bibitem{PS96} Petz D and Sud\'{a}r C 1996
{\sl J. Math. Phys.} {\bf 37} 2662

\bibitem{BC94} Braunstein S L and Caves C M 1994
{\sl Phys. Rev. Lett.} {\bf 72} 3439

\bibitem{Uh95} Uhlmann A 1995
{\sl Rep. Math. Phys.} {\bf 36} 461

\bibitem{Lu78} Lubkin E 1978
{\sl J. Math. Phys.}  {\bf 19} 1028  

\bibitem{LS88} Lloyd S and Pagels H 1988 
  {\sl Ann. Phys. (N.Y.)} {\bf 188} 186  

\bibitem{Pa93} Page D 1993
 {\sl Phys. Rev. Lett.} {\bf 71}  1291 

\bibitem{Me91}  Mehta M L 1991  {\sl Random Matrices} II ed.
 (New York: Academic)

\bibitem{Sl99b}  Slater P B 1999
{\sl J. Phys.} {\bf A 32} 8231

\bibitem{SZ03} Sommers H--J and {\.Z}yczkowski K 2003
{\sl J. Phys.} {\bf A 36} 10083 

\bibitem{ZS03}  {\.Z}yczkowski K and  Sommers H-J 2003 
{\sl J. Phys.} {\bf A 36} 10115 

\bibitem{SM99} Sengupta A M and  Mitra P P 1999
  {Phys. Rev.} {\bf E 60} 3389

\bibitem{pastur} Pastur L A 1972
{\sl Th. Math. Phys.} {\bf 10} 67 

\bibitem{FK94} Foong  S K and  Kanno S 1994
   {\sl Phys. Rev. Lett.} {\bf 72} 1148 

\bibitem{SR95}  S{\'a}nchez--Ruiz J 1995
{\sl Phys. Rev.} {\bf E 52} 5653 

\bibitem{Se96}  Sen S 1996
 {\sl Phys. Rev. Lett.} {\bf 77} 1

\bibitem{MML02} Malacarne L C and  Mandes R S  and Lenzi E K 2002
 {\sl Phys. Rev.} {\bf E 65} 046131 


\bibitem{NC00} Nielsen M A and Chuang I L 2000 
{\sl Quantum Computation and Quantum Information}
(Cambridge: Cambridge University Press)

\bibitem{itzykson} Itzykson C and Zuber J B 1980 {\sl J. Math. Phys.} {\bf 21}
411

\bibitem{harish} Chandra H 1957 {\sl Am. J. Math. } {\bf 79} 87
 
\end{thebibliography}
\end{document}